\documentclass{du-journals}
\journal{Journal of Holography Applications in Physics}
\ArticleType{Regular article}
\Year{Winter 2026}
\Vol{6}
\No{1}
\Page{\thepage--\pageref{LastPage}}
\Received{January xx, 2026}
\Revised{January xx, 2026}
\Accepted{January xx, 2026}
\Doihead{10.22128/jhap.2021.452.xxxx}
\title{Unified Entropic Dynamics Framework for Classical, and Quantum Wave Equations}

\author[1]{Shahid Nawaz}
\address[1]{Shaker High School, 445 Watervliet Shaker Rd, Latham, NY 12110, United States of America;\\ E-mail: snafridi@gmail.com}
\author[2]{Muhammad Saleem}
\address[2]{Department of Physics, Bellarmine University, 2001 Newburg Road, Louisville, KY 40205, United States of America;\\E-mail: msaleem@bellarmine.edu}
\author[3]{Muhammad S. Anwar}
\address[3]{Department of Engineering \& Construction, School of Architecture, Computing \& Engineering, University of East London, London E16 2RD, United Kingdom
and 
Department of Materials and Metallurgy, University of Cambridge, CB3 0FS Cambridge, United Kingdom;\\E-mail: m.s.anwar476@gmail.com}
\author[4]{Dalaver H.~Anjum}
\address[4]{Department of Physics, Khalifa University, Abu Dhabi, P. O. Box 127788, United Arab Emirates;\\ Corresponding Author E-mail: dalaver.anjum@ku.ac.ae}
\begin{document}

\begin{abstract}
Entropic Dynamics (ED) provides a statistical–inferential foundation for physical laws, deriving motion and field equations from principles of entropy maximization rather than quantization postulates. The ED reconstructs quantum mechanics by treating the evolution of probability distributions on configuration space as driven by information constraints, yielding the Schrödinger equation as a non-dissipative diffusion process. Building on this foundation, the present work extends the ED framework into a Unified Entropic Dynamics (UED) formulation that encompasses classical, quantum, relativistic, thermodynamic, and gravitational phenomena within a single information-geometric principle. By maximizing entropy subject to constraints on diffusion, drift, and gauge covariance over a manifold endowed with a supermetric $H_{ab}$, we derive a universal field equation that merges the Fokker–Planck and Hamilton–Jacobi structures into one covariant form. When specialized to different dynamical variables, this equation reproduces the simple harmonic oscillator, Schrödinger, Maxwell, Klein–Gordon, and gravitational wave equations, thereby revealing a deep equivalence between probabilistic inference and dynamical law. The UED framework demonstrates that spacetime geometry, quantum coherence, and thermodynamic diffusion emerge as complementary expressions of the same entropic process—establishing a unified inferential foundation for both microscopic and macroscopic physics. In this formulation, energy, probability, and entropy are intertwined aspects of information geometry, providing a consistent inferential foundation for understanding classical, quantum, and gravitational dynamics as complementary expressions of a single entropic law.
\end{abstract}

\newpage
\tableofcontents
\thispagestyle{dups}
\newpage

\section{Introduction}
One of the central challenges in modern theoretical physics is the reconciliation of classical mechanics, quantum mechanics, thermodynamics, and relativity within a single coherent framework~\cite{Jacobson1995, Silva2024, Dunkel2009, Hayward1997, Moradpour2024,faizal2025consequences}. Although each theory is remarkably successful in its own domain, they remain conceptually and mathematically distinct at both microscopic and macroscopic scales. Classical physics describes the deterministic motion of bodies through spacetime, quantum theory governs the probabilistic evolution of microscopic systems, thermodynamics captures macroscopic equilibrium and irreversible processes, and general relativity encodes gravity as spacetime curvature. Despite their apparent differences, all these domains share a common feature: the interplay between geometry, curvature, and information flow. In classical mechanics and wave theory, curvature links geometry to dynamics—the curvature of a potential surface defines restoring forces, stability, and oscillatory behavior. In quantum mechanics, it appears in the curvature of probability amplitudes within Hilbert or configuration space, determining interference and coherence. In thermodynamics, curvature arises in entropy landscapes that guide systems toward equilibrium. In general relativity, curvature becomes geometric in the most literal sense, describing how matter and energy determine the shape of spacetime itself. The laws of physics share an important conceptual and mathematical connection with B.~R.~Frieden’s Extreme Physical Information (EPI) principle ~\cite{FriedenSoffer1995}. The EPI approach regards physical laws as emergent from informational optimization. In that presented framework, physical laws arise from the extremization of the physical information functional. By using this approach, Frieden and Soffer showed that many fundamental equations of physics can be recovered from the condition of informational balance between data and source~\cite{FriedenSoffer1995}. In summary, it can be stated that the underlying variational principle operates in a space of probability amplitudes and treats Fisher information as the unifying geometric quantity that underlies both mechanics and field theory.

From the perspective of entropic dynamics (ED), these manifestations of curvature are not disparate but complementary: each expresses how information about the microscopic degrees of freedom of a system constrains its macroscopic evolution. Thus, the transition from energy curvature to entropic curvature represents a shift in the language of physics: from force-based determinism to inference-based dynamics. The ED framework, first introduced by Caticha et al.~\cite{Caticha2019EntropyQM,caticha2021entropic}, reconstructs physical laws not by quantizing pre-existing classical models but by applying the principles of entropic inference under appropriate constraints. By maximizing entropy with respect to a set of physically meaningful variables, ED derives both stochastic and deterministic evolution equations, recovering the Fokker--Planck and Hamilton--Jacobi forms as complementary limits of the same informational process. Their combination yields a generalized field equation structurally equivalent to familiar wave equations but derived from inference rather than postulation. This approach has been shown to reproduce nonrelativistic quantum mechanics, classical diffusion, and wave phenomena through purely statistical arguments~\cite{ipek2015entropic, Caticha2019EntropyQM, nawaz2012momentum}. In general, entropy and information-theory principles have emerged as powerful tools in rethinking the foundations of physics ~\cite{Caticha2019EntropyQM}. For example, modified theories of gravity have been derived from entropic considerations \cite{bianconi2025gravity}. Generally, the ED formulation seeks to reconstruct dynamical laws not by quantizing pre-existing classical systems, but by applying the rules of entropic inference under carefully chosen constraints. First developed in 2011~\cite{caticha2011entropic}, and successfully applied to a wide range of problems within nonrelativistic limits~\cite{ipek2015entropic,ipek2019entropic,ipek2020entropic,nawaz2016entropic,nawaz2024major,caticha2025entropic}. It has been demonstrated already that the ED formulation can be used to derive quantum mechanics as a form of inference driven by entropy maximization~\cite{ipek2015entropic,ipek2019entropic,ipek2020entropic,nawaz2016entropic,nawaz2024major}. Within this framework, two central ingredients are introduced: the entropy functional and the energy functional. The entropy functional, once constrained appropriately, leads to the Fokker-Planck (FP) equation, whereas the energy functional produces the Hamilton–Jacobi (HJ) equation. Combining these yields the Schrödinger equation, thereby recovering the structure of non-relativistic quantum mechanics from the principles of inference rather than postulation. Non-relativistic entropic dynamics (NED) assumes that a particle inhabits a three-dimensional (3D) configuration space, where the fundamental object of the study is the transition probability $P(x'|x)$ between successive positions. Because this probability is timeless, an additional construct- entropic time - is introduced to order sequences of transitions. Relativistic entropic dynamics (RED), by contrast, supposedly embeds the particle in a four-dimensional spacetime, with time incorporated directly into the transition probability. In this way, the generated formulation should result in dynamical equations that closely resemble the Klein-Gordon equation, thereby linking ED to relativistic quantum mechanics in a natural manner.
The present work extends the ED framework as a unified entropic dynamics (UED) to encompass confined or correlated relativistic and gravitational domains, thereby establishing a unified geometric foundation for classical, quantum, thermodynamic, and spacetime dynamics. The resulting generalized field equation, which has been derived from entropy maximization, governs the evolution of arbitrary dynamical variables $h_a$ on an information-theoretic manifold endowed with a supermetric $H_{ab}$. When specialized to different choices of $h_a$, in the generalized equation, it yields the fundamental laws of physics as limiting cases: the harmonic oscillator, Maxwell’s electromagnetic equation, the classical heat equation, the Schr\"odinger equation, and the relativistic Klein-Gordon equation. A key advancement reported here is the covariant extension of ED to curved spacetime. By replacing it with $H^{ab}$ with the spacetime metric $g^{\mu\nu}$ and introducing covariant derivatives, the ED equation becomes a geometric wave equation whose linearized form reproduces the standard gravitational wave equation in the transverse traceless gauge (TT). In this interpretation, gravitational waves correspond to oscillations of the spacetime information metric—a macroscopic manifestation of the same entropic principles that govern microscopic quantum and thermodynamic processes. This unification suggests that all forms of dynamical evolution, from quantum coherence to spacetime curvature, arise from a single informational principle: the maximization of entropy under geometric and physical constraints. The remainder of this paper is structured as follows. Section~2 develops the UED formulation and derives the unified entropic dynamics equation from maximization of entropy with diffusion, drift, and gauge covariance constraints. Section~3 applies the equation to the scalar case, reproducing simple harmonic motion as the entropic analog of Newtonian mechanics. Section~4 treats the three-vector case, recovering Maxwell's, Fourier's, and Schrodinger's equations as distinct limits of the same informational structure. Section~5 extends the UED framework to the four-vector case, deriving both the relativistic Klein--Gordon and the gravitational-wave equations from entropic inference. Section~6 discusses the broader implications of these results, highlighting how energy, probability, and entropy emerge as interwoven aspects of information geometry.  


\section{Derivation of the Generalized Entropic Dynamics Framework}\label{sec-ED}
Here we develop a general entropic dynamics or the UED construction valid for any dynamical variable \(h_{a\ldots}\) living on a \(d\)-dimensional configuration manifold with supermetric \(H_{ab\ldots}\). Our goal is the short-step transition probability \(P(h'|h)\), obtained by maximizing the relative entropy;

\begin{equation}
S[P,Q] \;=\; -\!\int dh'\, P(h'|h)\,
\log\!\left(\frac{P(h'|h)}{Q(h'|h)}\right)
\label{eqn:Spq}
\end{equation}

with respect to covariant ignorance prior \(Q(h'|h)\propto H^{1/2}(h')\), subject to physically motivated constraints: (i) a fixed mean squared step \(\langle \Delta h_{a\ldots}\Delta h_{b\ldots}\rangle\) (diffusive motion), (ii) a drift driven by a potential \(\varphi\), and (iii) gauge-covariant coupling to a tensor potential \(A_{a\ldots}\), together with normalization. The resulting variational problem yields a Gaussian kernel whose mean and covariance are set by Lagrange multipliers; iterating short steps produces a Fokker-Planck equation, and adding an energy functional gives a Hamilton-Jacobi companion. Next, we specify the prior, impose relevant constraints, and maximize the entropy functional. The prior probability codifies any relation between $h$ and $h^\prime$ before processing the actual information contained in the relevant constraints. Assign equal probabilities to equal volumes. That is, in the case of ignorance, the prior is uniform.
\begin{equation}
    Q(h^\prime|h)\propto H^{\frac{1}{2}}(h^\prime)
\label{Q-d}
\end{equation}

Here, \(Q(h'|h)\) encodes our state of knowledge \emph{prior} to processing the constraints. To make 'ignorance' precise in an arbitrary configuration space \(M\), let \(h^A\) (a multiindex collecting \(a\ldots\)) be a local dynamical coordinate, and let the positive-definite supermetric \(H_{ab}(h)\) endow \(M\) with Riemannian volume. It is equal to a  dyadic of the basis vectors of $h_a$ and $h_{b}$ dynamical variables of form $H_{ab}=e_{a}\otimes e_{b}$. Where $e_{a}$ and $e_{b}$ are base vectors of $h_{a}$ and $h_{b}$, respectively. A prior that is genuinely uninformative must be invariant under reparameterizations \(h\mapsto \tilde h(h)\). Writing \(J^A{}_B=\partial \tilde h^A/\partial h^B\), the metric transforms as \(H\mapsto \tilde H=J^{\!\top}HJ\)so:
\begin{equation}
 \sqrt{\det \tilde H(\tilde h)}\,d^N\tilde h =\sqrt{\det H(h)}\,d^N h   
\end{equation}.
Requiring the prior density to define an invariant measure \(Q(h'|h)\,d^N h'\) then singles out the Jeffreys prior on \(M\) in its uniform form ~\cite{Jaynes2003}, i.e.,
\begin{equation}
Q(h'|h)=\frac{1}{Z(h)}\,\sqrt{\det H(h')}\,,
\end{equation}
with \(Z(h)\) a local normalizer (for example, over a small geodesic ball when generating short-step kernels). Herein $Z(h)$ is taken as equal to one. Physically, \(H_{ab}\) is the natural 'kinetic' or information metric in the space of dynamical variables: in mechanics it coincides with the metric from the quadratic kinetic term, in quantum–gravitational applications with the DeWitt supermetric on superspace, and within the ED it plays the role of the Fisher–information metric induced by constraints. This choice guarantees coordinate-free inference and ensures that the ensuing maximum-entropy update produces a Gaussian short–step kernel whose covariance is \(H^{-1}\), which will underlie the Fokker-Planck/Hamilton-Jacobi structure derived next.
 After knowing the prior, we specify the constraints. The first constraint concerns the transition probability $ P(h^\prime|h)$. It deals with short steps (infinitesimal). A large step (finite) can be obtained by accumulating short steps
\begin{equation}
    \int dh^\prime P(h^\prime|h)H_{ab\ldots}\Delta h^{a\ldots}\Delta h^{b\ldots}=\kappa_s
\label{eqn:dh}
\end{equation}
where $\kappa_s$ is a small constant. The second constraint concerns the draft potential, or
\begin{equation}
    \int dh^\prime P(h^\prime|h)\Delta h^{a\ldots}\frac{\partial\phi}{\partial h^{a\ldots}}=\kappa_f
\label{eqn:dhP}
\end{equation}
where $\phi$ is the drift potential, which turns out to be the phase of the wave function, and $\kappa_f$ is a constant. The third constraint takes into account the gauge invariance.
\begin{equation}
\int dh^\prime P(h^\prime|h)\Delta h^{a\ldots} \mathcal{A}_{a\ldots}=\kappa_A   
\end{equation}
 where $\mathcal{A}_{a\ldots}$ is the gauge-invariantiant tensor potential and $\kappa_A$ is a constant. The last constraint is normalization. Finally, incorporate the prior and the constraints and maximize the entropy functional. The result is    
\begin{equation}
    P(h^\prime|h)=\frac{H^{1/2}(h^\prime)}{\zeta(h)}\exp\left[-\frac{1}{2\sigma^2}H_{ab\ldots}(\Delta h^{a\ldots}-\langle\Delta h^{a\ldots}\rangle)(\Delta h^{b\ldots}-\langle\Delta h^{a\ldots}\rangle) \right]
\label{P-d}
\end{equation}
where the expectation is given by
\begin{equation}
    \langle\Delta h^{a\ldots}\rangle=\sigma^2H^{ab\ldots}\left(\alpha\frac{\partial\phi}{\partial h^{a\ldots} }-\beta \mathcal{A}_{a\ldots}\right)
\label{h-d}
\end{equation}
where $\sigma^2, \alpha$, and $\beta$ are Lagrange multipliers, and
\begin{equation}
    \langle\Delta h^{a\ldots}\Delta h^{b\ldots}\rangle=\sigma^2 H^{ab\ldots}
\label{eqn:deltaH}
\end{equation}
The process is driven by Brownian motion because $\Delta h^{a\ldots}\sim\sqrt{\sigma^2}$. The Brownian motion leads to the Fokker-Planck equation as is in Eq.~\ref{eqn:deltaH}
\begin{equation}
    \frac{1}{\sqrt{H}}\frac{\partial}{\partial h^{a\ldots}}\left(\sqrt{H}H^{ab\ldots}\rho\left(\frac{\partial\varphi}{\partial h^{b\ldots}}-\beta\mathcal{A}_{b\ldots}\right)\right)=0
\label{eq:FP}
\end{equation}
where $\rho(h)$ is the marginalized probability distribution and 
\begin{equation}
    \varphi=\alpha\phi-\left(\log\rho\right)^{\frac{1}{2}}
\label{eq:MargPDF}
\end{equation}
$\rho$ and $\varphi$ are dynamical variables. This is the case if, in addition to Eq.~\ref{eq:FP}, we impose a second constraint.
\begin{equation}
    \mathcal{H}=\int dh\sqrt{H} \rho \left(\frac{1}{2}H^{ab\ldots}\left(\frac{\partial\varphi}{\partial h^{a\ldots}}-\beta\mathcal{A}_{a\ldots}\right)\left(\frac{\partial\varphi}{\partial h^{b\ldots}}-\beta\mathcal{A}_{b\ldots}\right)+\frac{1}{4\rho^2}H^{ab\ldots}\frac{\partial\rho}{\partial h^{a\ldots}}\frac{\partial\rho}{\partial h^{b\ldots}}+\mathcal{M}\right)
\label{eq:H-after2ndConstr}
\end{equation}
 Where $\mathcal{M}$ is independent of $\rho$ and $\varphi$. It can be thought of as entropic mass that balances or acts as a source term during the evolution of entropy under dynamical variables. Once the dynamical variable $h^{a\ldots}$ is set or known, then the $\mathcal{M}$ can be chosen according to dimension analysis by setting it proportional to the inner product of the conjugate variable of the dynamical variable. The minimization of $H$ as  a function of $\rho$ and $\varphi$ needs to be done, and the following relation can be obtained by doing so:  
\begin{equation}
  H^{ab\ldots}\left(\frac{\partial\varphi}{\partial h^{a\ldots}}-\beta \mathcal{A}_{a\ldots}\right)\left(\frac{\partial\varphi}{\partial h^{b\ldots}}-\beta \mathcal{A}_{b\ldots}\right)+\mathcal{M}-\frac{\Delta_H(\sqrt{\rho})}{\sqrt{\rho}}=0
\label{eq:H-min}
\end{equation}
where
\begin{equation}
 \frac{\Delta_H(\sqrt{\rho})}{\sqrt{\rho}}=\frac{1}{2\rho\sqrt{H}}\frac{\partial}{\partial h^{a\ldots}}\left(\sqrt{H}H^{ab\ldots}\frac{\partial\rho}{\partial h^{b\ldots}}\right) -\frac{1}{4\rho^2}H^{ab\ldots}\frac{\partial\rho}{\partial h^{a\ldots}}\frac{\partial\rho}{\partial h^{b\ldots}} 
\label{HJ} 
\end{equation}
Eqs.~\ref{eq:FP} and~\ref{HJ} can be combined into a single equation by using
\begin{equation}
    \Psi=\sqrt{\rho}e^{i\phi}
\end{equation}
We obtain
\begin{equation}
 \frac{1}{\sqrt{H}}\left(i\frac{\partial}{\partial h^{a\ldots}}
 -\beta \mathcal{A}_{a\ldots}\right)\sqrt{H}H^{ab\ldots}\left(i\frac{\partial}{\partial h^{b\ldots}}-\beta\mathcal{A}_{b\ldots}\right)\Psi+\mathcal{M}\Psi=0
\label{SE-d}   
\end{equation}

Eq.~\ref{SE-d} represents the generalized framework or the UED framework sought to describe the wave dynamics of particles and fields $(\Psi)$ in an arbitrary potential $\mathcal M$ depending on the configuration space. In quantum theory, particles obey the rules of first quantization. In contrast, the fields follow the rules of second quantization. The inner product of conjugate variables will turn out to be a scalar invariant. Starting from a maximization of entropy with carefully chosen priors and constraints, it yields a transition probability that resembles Brownian motion and naturally leads to the Fokker–Planck and Hamilton–Jacobi equations. Introducing the wave function $\Psi=\sqrt{\rho}e^{i\phi}$, these ingredients combine into a single compact form, Eq.~\ref{SE-d}. This equation is a generalized field equation where the operator acting on $\Psi$ involves derivatives with respect to the dynamical variables, modified by a vector potential $\mathcal{A}_\mu$ and supplemented by a mass-like term $\mathcal{M}$. When a specific form of the dynamical variables is considered, the unified entropic dynamics equation yields their corresponding field equations, each with a specific interpretation of $\mathcal{M}$ fixed by dimensional analysis. The product of these conjugate variables naturally forms a scalar quantity that appears in the mass-like term $\mathcal{M} = -p_a p^a / \hbar^2$, linking the dynamical structure of the system to its geometric properties, but is derived here from purely entropic principles rather than postulated directly. The significance of Eq.~\ref{SE-d} is that it provides a unifying template capable of describing a wide class of physical systems depending on the choice of dynamical variable $h^{a\ldots}$. When $h^{a\ldots}$ is taken to be spacetime events $x^{\mu}$, Eq.~\ref{SE-d} reduces to the Klein–Gordon equation, demonstrating consistency with relativistic quantum theory. Therefore, the Eq.~\ref{SE-d} encapsulates a general dynamical law from which many known equations of physics can be derived, supporting the idea that entropic dynamics may serve as a foundational framework that bridges statistical inference with quantum and gravitational theories. 


\section{Case-I: Dynamical Variable As a time Scalar}\label{sec-SHO}
For a case in which the generalized dynamical variable $h^a$ introduced in Eq.~\ref{HJ} is chosen to be a \textit{scalar quantity}, both $\sqrt{H}$ and $H^{ab\ldots}$ reduce to unity. The generalized field equation further simplifies considerably if the particle dynamics is considered in the absence of external fields, i.e. $\mathcal{A}=0$. As a result, there are no tensor indices or vector potentials to account for, and the system represents the simplest possible configuration space-one described by a single coordinate $h$ and its conjugate variable. This scalar case serves as a pedagogical example showing that the entropic dynamics framework naturally reproduces classical oscillatory motion. Starting from the unified entropic dynamics equation,
\begin{equation}
\left[-\frac{\partial^2}{\partial h^2} +\mathcal{M}\right]\Psi = 0,
\label{eq:mastary}
\end{equation}
We consider the situation in which the potential term vanishes and the system evolves freely in the scalar coordinate $h$. 
Here, the differential operator $-\partial^2/\partial h^2$ plays the role of a kinetic operator, analogous to the Laplacian in wave mechanics, while the term $\mathcal{M}$ acts as an effective mass or energy constant that characterizes the curvature of the potential landscape in configuration space. 
Eq.~\ref{eq:mastary} is thus the entropic analog of a one-dimensional Helmholtz equation, describing standing or propagating modes depending on the sign of $\mathcal{M}$. In the scalar-time realization of the unified entropic dynamics framework, the quantity $\mathcal{M}$ acquires a clear interpretation as an information mass, encoding the intrinsic energetic content associated with temporal oscillations. When the dynamical variable is chosen to be the time scalar $h = t$, its conjugate variable is the angular frequency $\omega$, which characterizes the rate of phase accumulation in time. Dimensional consistency and entropic closure then require the information mass term to be proportional to the square of this conjugate quantity, $\mathcal{M} \propto \omega^{2}$, equivalently proportional to the square of the system’s energy scale. In classical mechanics, this identification is naturally realized in the simple harmonic oscillator, where the equation of motion $\ddot{x} + \omega^{2} x = 0$ describes bounded, periodic evolution under a restoring force. Within the entropic dynamics formulation, the same structure emerges from the scalar UED equation $(-\partial_{t}^{2} + \mathcal{M})\Psi = 0$, whose oscillatory solutions demand $\mathcal{M} = -\omega^{2}$. Thus, the information mass $\mathcal{M}$ plays the role of an entropic restoring parameter, encoding the curvature of the informational landscape in time and ensuring coherent oscillatory dynamics. This establishes a direct correspondence between classical mechanical stability and informational curvature, showing that the squared frequency---or equivalently the squared energy scale---acts as the fundamental scalar invariant governing temporal dynamics in the entropic framework. Therefore, the equation~\ref{SE-d} is the canonical form of the simple harmonic oscillator in the representation of entropic dynamics. If the scalar variable $h$ is set as the physical time coordinate, $h = t$, then the conjugate variable of $t$ is angular frequency $\omega$. Consequently, $\Psi$ representing the displacement $(\mathbf{X})$ of the particle in harmonic motion and therefore reducing Eq.~\eqref{eq:mastary} or, in turn, Eq.~\ref{SE-d} to the form given below:
\begin{equation}
\left[\frac{\partial^2}{\partial t^2} + \omega^2\right] \mathbf{X(t)} = 0    
\label{eq:HM-X}
\end{equation}

Eq.~\ref{eq:HM-X} is precisely the differential equation that governs the temporal harmonic oscillations. Its general solution $\mathbf{X}(t) = A e^{i\omega t} + B e^{-i\omega t}$ describes sinusoidal oscillations of amplitude determined by the constants $A$ and $B$, corresponding to time-propagating modes forward and backward. Physically, this result demonstrates that simple harmonic motion, a cornerstone of classical and quantum physics, emerges directly from the formalism of entropic dynamics when the system's configuration space is one-dimensional. The oscillations reflect the system’s attempt to restore maximum entropy under the constraint of finite energy, thereby linking information-theoretic inference to the familiar restoring forces of Newtonian mechanics. In this sense, Eq.~\ref{eq:mastary} represents not only the classical harmonic oscillator but also the fundamental time-dependent mode of all wavelike phenomena within the entropic framework. This equation can also be shown to be mathematically identical to the standard form of the simple harmonic oscillator. Interpreting $\mathbf{X}$ as the generalized coordinate (or, in a quantum setting, using Ehrenfest’s theorem to identify $\langle x \rangle$ with the classical trajectory) recovers Newton’s law with a linear restoring force. Near any stable equilibrium, the potential energy $V(x)$ becomes $V(x) \approx \tfrac{1}{2} k x^{2}$, therefore, the mapping of $\omega^{2}$ becomes $\omega^{2} = k/m$. In summary, Equation~\ref{eq:mastary} is exactly Newton’s second law for a linear restoring force, with $\omega^{2} = k/m$; it can be obtained by force balance, by the Lagrangian, or by your reduction in entropic dynamics $\mathcal{M} = -\omega^{2}$ that yields the same simple harmonic oscillator (SHO) dynamics.


\section{Case-II: Dynamical Variable As a Spatial Three-Vector}
In the preceding section, the generalized entropic dynamics framework was applied to a scalar dynamical variable to demonstrate how simple harmonic motion –and by extension Newton's second law –emerges naturally from the maximization of entropy subject to energy constraints. The scalar case served as the simplest realization of entropic inference in the configuration space, where the oscillatory behavior of the system was encoded in the curvature of its information potential. We now extend this reasoning to a richer configuration space in which the dynamical variable $h_a$ is promoted to a three-vector. This extension allows the formalism to describe both classical fields and particle wavefunctions that depend on spatial coordinates, thereby linking the statistical structure of entropic dynamics to the geometry of three-dimensional space.

When $h_a$ is taken as the spatial vector $x_i$, the generalized field equation~\ref{SE-d} encompasses three distinct but complementary physical regimes that emerge from the same inferential structure. In the classical limit, it reduces to the Maxwell wave equation, which describes the propagation of electromagnetic disturbances as an optimal flow of information through spacetime~\cite{MIT2024}. When the entropic mass term is reinterpreted as a generator of temporal diffusion, the same framework produces the heat equation~\cite{Onsager2024, Thermal2025}, showing that thermal conduction and energy dissipation arise as entropic processes governed by the geometry of information. In the quantum-mechanical limit, the equation reduces to the Schrödinger equation, revealing that quantum behavior arises from the same inference principles that govern classical and thermodynamic evolution. This tripartite correspondence demonstrates the universality of the entropic dynamics approach: whether describing electromagnetic radiation, heat diffusion, or matter waves, the theory unifies their evolution under a single informational law. The subsections that follow develop these three cases in detail: Beginning with the electromagnetic formulation, extending to the thermodynamic interpretation, and concluding with the non-relativistic quantum description.

\subsection{Electromagnetic Wave Equation}
The electromagnetic field represents one of the most fundamental realizations of wave phenomena in nature.  As mentioned earlier, within the entropic dynamics (ED) framework, electromagnetic waves will emerge not merely as solutions to Maxwell’s equations but as natural outcomes of entropy maximization under geometric and gauge constraints. We start by setting $h$ to a position vector $(\mathbf{x})$, then $\Psi$ becomes a function of $\mathbf{x}$ that can also evolve in time $t$. This makes $\Psi$ as $\Psi(\mathbf{x}, t)$ in eq.~\eqref{SE-d}, which takes the form shown below:
\begin{equation}
    \left(-\mathbf{\nabla}^{2}+\mathcal{M}\right)\Psi(\mathbf{x}, t)=0
\label{eq:delsquare}
\end{equation}

The quantity $\mathcal{M}$ will be proportional to the conjugate variable of $\mathbf{x}$, which is the linear momentum $(\mathbf{p})$ in this case. Here, $\nabla^2$ acts as the spatial curvature operator. In ED, the term $\mathcal{M}$ plays a role similar to that of a mass or potential operator, determining how wave $\Psi(\mathbf{x}, t)$ oscillates in space and time. Thus, the quantity $\mathcal{M}$ can be written as
\begin{equation}
\mathcal{M} \propto p^{2}.
\end{equation}
For an oscillating classical wave characterized by angular frequency $\omega$ and wave number $k = |\mathbf{k}|$, the dispersion relation for a free wave is
\begin{equation}
p = \hbar k, \qquad \omega = c k.
\end{equation}
Hence, for a unit proportionality constant,
\begin{equation}
\mathcal{M} = k^{2} = \frac{\omega^{2}}{c^{2}}.
\end{equation}

The value of $\mathcal{M}$ as $\frac{\omega^{2}}{c^{2}}$ in eq~\ref{eq:delsquare} turns the classical wave  $\Psi(\mathbf{x}, t)$ into an oscillating wave whose phase varies as $\mathbf{k}.\mathbf{x}-\omega t$. This means we can equate the quantity $\mathcal{M}$ with $\frac{1}{c^{2}}\frac{\partial^2}{\partial t^2}$. Therefore, the~\ref{eq:delsquare} can be written in the following form:
\begin{equation}
    \left(\mathbf{\nabla}^{2}-\frac{1}{c^{2}}\frac{\partial^2}{\partial t^2}\right)\Psi (\mathbf{x}, t)=0
    \label{eq_4.5}
\end{equation}
The above equation is Maxwell's wave equation that arises naturally from the equation ~\eqref{SE-d} in free space. It describes how electromagnetic waves, such as light, propagate through space and time. Here, $\nabla^{2}$ represents the Laplacian operator, accounting for the spatial variation of the wave, while $\frac{1}{c^2}\frac{\partial^{2}}{\partial t^{2}}$ represents its temporal variation. The function $\Psi(\mathbf{x},t)$ can represent either the electric field $\mathbf{E}$ or the magnetic field $\mathbf{B}$. This equation shows that both fields satisfy the same fundamental wave equation, meaning they propagate as light in a vacuum. It encapsulates the idea that disturbances in the electromagnetic field spread throughout space without requiring a medium, which was a revolutionary insight of Maxwell’s theory. Equation~\eqref{SE-d} in the entropic dynamics framework is not just a mathematical generalization of familiar wave equations; it represents a deeper statistical foundation for physics, where the laws of dynamics are derived from principles of inference and maximization of entropy. When this equation is specialized to the case of electromagnetic waves, it naturally reduces to Maxwell’s wave equation \eqref{eq_4.5}, showing that what we typically treat as a postulate of field theory can instead be viewed as an emergent property of information-theoretic constraints~\cite{nawaz2024major}. In this sense, the entropy-based approach connects the microscopic statistical structure of field fluctuations with macroscopic wave propagation. The role of entropy here is to encode uncertainty in the field configurations, and by maximizing entropy subject to constraints (such as gauge covariance and energy conservation), the resulting dynamics enforce wave-like behavior. This provides a deeper physical picture: electromagnetic radiation can be understood not only as solutions to Maxwell’s equations but as manifestations of an underlying entropic process, where the wave equation reflects the optimal flow of information consistent with the symmetries of spacetime. Thus, the link between Eq.~\eqref{SE-d} and Maxwell’s wave equation highlights how entropy and information theory can serve not only as the ground for classical wave theory but also for classical field theory.

\subsection{Heat Wave Equation}
Heat conduction bridges microscopic fluctuations and macroscopic energy transport. In contrast to purely oscillatory dynamics, thermal diffusion embodies the irreversible flow of information toward equilibrium. Within the entropic dynamics (ED) framework, it will be shown below that this process emerges naturally when the entropic mass term is reinterpreted as a generator of temporal relaxation rather than wave propagation. We start from the generalized ED field equation, i.e., Eq.~\ref{SE-d}, on a configuration manifold with a metric $H_{ab}$ that can be mathematically expressed in the following way,
\begin{equation}
\partial_a\!\left(H^{ab}\,\partial_b \Psi\right)+\mathcal{M}\,\Psi=0,
\label{eq:master}
\end{equation}
and specialize to the three-vector case $h_a=x_i$, so that $a,b\in\{1,2,3\}$ and $\Psi\equiv \Psi(\mathbf{x},t)$. To model thermal diffusion, we identify the field with temperature, $\Psi(\mathbf{x},t)\equiv T(\mathbf{x},t)$, and promote the scalar term $\mathcal{M}$ to the time evolution generator of a parabolic dynamic. A dimensionally consistent closure is~\cite{Bird2007, Pope2000, Kays1993, DeGroot1984};
\begin{equation}
\mathcal{M}\,\Psi\;\;\longrightarrow\;\;-\frac{1}{\alpha}\,\partial_t \Psi,
\label{eq:closure}
\end{equation}
where $\alpha$ has units of diffusivity (m$^2$/s). Substituting \eqref{eq:closure} into \eqref{eq:master} yields
\begin{equation}
\partial_a\!\left(H^{ab}\,\partial_b T\right)=\frac{1}{\alpha}\,\partial_t T.
\end{equation}
Rearranging gives the tensorial heat equation in the following form,
\begin{equation}
\;\partial_t T \;=\; \alpha\,\partial_i\!\left(H^{ij}\,\partial_j T\right)\;,
\label{eq:heat-tensor}
\end{equation}
which is the anisotropic diffusion law on the configuration manifold with metric $H^{ij}$. In continuum thermodynamics, Fourier law with anisotropic conductivity $k^{ij}$ and energy balance implies
\begin{equation}
\rho c\,\partial_t T \;=\; \partial_i\!\left(k^{ij}\,\partial_j T\right).
\label{eq:fourier}
\end{equation}
Comparing \eqref{eq:heat-tensor} and \eqref{eq:fourier} identifies the geometric metric with the transport tensor via
\begin{equation}
k^{ij}\;=\;\rho c\,\alpha\,H^{ij},
\qquad\text{so that}\qquad
\alpha\,H^{ij}=\frac{k^{ij}}{\rho c}.
\label{eq:ident}
\end{equation}
Thus, the information-geometry metric $H^{ij}$ encodes the directional ease of heat flow (anisotropy) and reduces to the identity in homogeneous isotropic media. For the case of the isotropic limit, $H^{ij}=\delta^{ij}$,
\begin{equation}
\;\partial_t T \;=\; \alpha\,\nabla^2 T\;,
\label{eq:heat-iso}
\end{equation}
with $\alpha=k/(\rho c)$ the standard thermal diffusivity. A plane wave ansatz $T\sim e^{i(\mathbf{k}\cdot\mathbf{x}-\omega t)}$ in \eqref{eq:heat-tensor} can be chosen that gives the following dispersion relation,
\begin{equation}
-i\omega \;=\; -\alpha\,k_i H^{ij}k_j,
\end{equation}
so $\omega=i\,\alpha\,k_i H^{ij}k_j$ has purely imaginary frequency, as required for diffusive (non-oscillatory) decay. In the isotropic case $H^{ij}=\delta^{ij}$ this reduces to $\omega=i\,\alpha |\mathbf{k}|^2$. The equation~\eqref{eq:heat-tensor} admits the usual boundary conditions: Dirichlet $T|_{\partial\Omega}=T_b$, Neumann $n_i k^{ij}\partial_j T|_{\partial\Omega}=q_n$, or Robin $n_i k^{ij}\partial_j T + h(T-T_\infty)=0$, with $k^{ij}$ related to $H^{ij}$ by \eqref{eq:ident}~\cite{Kreyszig2011}.

\subsection{Nonrelativistic Quantum Mechanical Wave Equation}
Quantum mechanics represents the most profound expression of wave behavior in nature, where probability, rather than energy or temperature, becomes the fundamental carrier of information. Within the entropic dynamics (ED) framework, it will be shown herein that the emergence of quantum laws follows directly from the same inferential principles that generate classical and thermodynamic equations. Setting the dynamical variable $h_i$ the same as the spatial coordinates $x_i$, it can be shown that Equation~\ref{SE-d} of the entropic dynamics framework directly connects to the foundations of quantum mechanics. In this context, it $h_i = x_i$ describes a particle moving in three-dimensional Euclidean space, where the underlying metric is $\eta_{ij} = \delta_{ij}$. The generalized equation~\ref{SE-d} therefore reduces to a purely spatial form involving the Laplacian operator, which governs the curvature of the probability amplitude in configuration space. Starting from the simplified form,
\begin{equation}
(-\nabla^2 + \mathcal{M})\Psi = 0\label{34}
\end{equation}
We interpret the term $-\nabla^2$ as the spatial kinetic contribution, reflecting how the wavefunction $\Psi$ varies in space, while the scalar quantity $\mathcal{M}$ encapsulates an intrinsic energy-like or mass-like contribution determined by the system’s dynamics. Equation~\ref{34} has the mathematical structure of the Helmholtz equation, which in wave theory describes standing waves in free space or confined regions. In the ED framework, this represents a balance between the diffusive spread of the probability distribution and the constraint imposed by the energy functional. To explore the wave nature of this solution, we assume a plane-wave ansatz:
\begin{equation}
\Psi = e^{\frac{i}{\hbar} \mathbf{p} \cdot \mathbf{x}}\label{35}
\end{equation}
where $\mathbf{p}$ is the conjugate momentum associated with the spatial coordinate $\mathbf{x}$, and $\hbar$ is the reduced Planck constant. This trial solution embodies the fundamental Fourier duality between position and momentum spaces, a central concept in quantum theory. It indicates that momentum arises as the conjugate variable to position in the entropic configuration manifold. Substituting Equation~\ref{35} into ~\ref{34} yields
\begin{equation}
 (-\nabla^2)\Psi = \frac{p^2}{\hbar^2}\Psi\label{36}   
\end{equation}

which leads to the identification
\begin{equation}
\mathcal{M} = -\frac{p^2}{\hbar^2}
\label{eq:M}
\end{equation}
This relationship reveals that $\mathcal{M}$ acts as a bridge between the geometric operator $-\nabla^2$ and the physical momentum $\mathbf{p}$. The negative sign indicates oscillatory or wave-like behavior, analogous to how the curvature of $\Psi$ in configuration space corresponds to momentum and kinetic energy in physical space. The total energy of a non-relativistic particle is given by
\begin{equation}
E = \frac{p^2}{2m} + V\label{39}
\end{equation}
where the first term represents kinetic energy and $V$ is the potential energy. Rearranging this expression gives
\begin{equation}
p^2 = 2m(E - V) \label{eq:psquare}
\end{equation}
This substitution links the momentum appearing in the entropic variable $\mathcal{M}$ to the observable energy and potential terms in classical mechanics. Substituting Eq.~\ref{eq:psquare} into Eq.~\ref{eq:M} produces
\begin{equation}
\mathcal{M} = -\frac{2m}{\hbar^2}(E - V)\label{40}
\end{equation}
This step is critical-it translates the statistical term $\mathcal{M}$, originally introduced through entropic constraints, into the familiar energy difference $(E - V)$. Thus, the quantity $\mathcal{M}$ encodes both quantum and classical energy content in a unified expression. By substituting Equation~\ref{40} back into the wave equation~\ref{36}, we obtain
\begin{equation}
E\Psi = \left(-\frac{\hbar^2}{2m}\nabla^2 + V\right)\Psi.
\end{equation}
This expression recovers the time-independent Schrödinger equation, showing that quantum mechanics arises not as a postulate but as a logical consequence of the entropic dynamics framework. Here, the Laplacian term represents diffusion-like propagation in configuration space constrained by the potential energy landscape, while the eigenvalue $E$ quantifies the energy levels consistent with the entropic equilibrium. To extend this to the time-dependent form, we recall that energy acts as the generator of temporal evolution. In quantum theory, this is expressed through the operator substitution $E \rightarrow i\hbar \frac{\partial}{\partial t}$. Applying this to Equation~\ref{39} yields
\begin{equation}
i\hbar\frac{\partial\Psi}{\partial t} = \left(-\frac{\hbar^2}{2m}\nabla^2 + V\right)\Psi.
\end{equation}
Equation~\ref{40} is the full time-dependent Schrödinger equation, the cornerstone of non-relativistic quantum mechanics. Physically, it describes how the wavefunction $\Psi(\mathbf{x},t)$ evolves under both spatial and temporal constraints, continuously updating the probability distribution in response to information encoded in the potential $V(\mathbf {x})$. In the entropic dynamics picture, this equation has deep interpretational significance: time evolution arises from sequential entropy updates rather than from an external clock, and the Laplacian term corresponds to information diffusion driven by uncertainty in position. Thus, the Schrödinger equation emerges as an optimal inference law---one that maximizes entropy while conserving energy and maintaining gauge consistency. This reformulation shows that quantum dynamics can be understood as a special case of information-based reasoning applied to physical systems.

\section{Case-III: Dynamical Variable As a spacetime Four-Vector}\label{sec-KGE}
The preceding sections established that the entropic dynamics (ED) framework successfully reproduces classical, quantum, and thermodynamic laws as particular realizations of an underlying inferential principle. Having demonstrated that the same information-geometric foundation yields the harmonic oscillator, Maxwell’s electromagnetic equations, the classical heat equation, and the Schr\"odinger equation, we now extend the formalism to the relativistic domain. In this section, the generalized field equation is applied to the case where the dynamical variable $h_\mu$ corresponds to spacetime events $x_\mu$. This identification embeds the ED framework directly within the four-dimensional structure of special relativity, allowing us to investigate how inference-based dynamics reproduce the fundamental equations governing relativistic quantum fields and spacetime curvature.

We begin by considering a flat Minkowski background, where $H_{\mu\nu}$ reduces to $\eta_{\mu\nu}$, and show that the generalized ED equation naturally transforms into the Klein--Gordon equation. This demonstrates that relativistic quantum mechanics, often introduced as an independent postulate, can instead be derived from entropic inference and information geometry. The analysis is then extended to curved spacetime by replacing $\eta_{\mu\nu}$ with a general metric $g_{\mu\nu}(x)$ and substituting ordinary derivatives with covariant derivatives. This covariant generalization leads to the curved-space ED equation, which reduces, in the linearized limit, to the standard gravitational-wave equation in the transverse--traceless (TT) gauge. Thus, Section~5 unifies the relativistic and gravitational domains within the same entropic framework, revealing that both matter waves and spacetime perturbations emerge as expressions of information flow through curved geometric manifolds.

\subsection{Relativistic Quantum Mechanical Wave Equation}
We show here that the relativistic extension of the entropic dynamics (ED) framework provides a natural bridge between probabilistic inference and spacetime geometry. In this domain, the dynamical variable is elevated from spatial coordinates to four-dimensional spacetime events, allowing the inferential structure of ED to incorporate Lorentz invariance. The resulting formulation reveals that relativistic wave behavior---long considered a separate postulate of quantum field theory---emerges directly from the maximization of entropy under covariant constraints. By applying the generalized ED field equation in a flat Minkowski background, one obtains a hyperbolic wave equation whose structure and invariance properties coincide with those of the Klein--Gordon equation. This demonstrates that the relativistic energy--momentum relation, $E^{2} = p^{2}c^{2} + m^{2}c^{4}$, arises as an informational identity connecting energy, curvature, and entropy within the geometry of inference. Hence, the ED framework not only reproduces the formalism of relativistic quantum mechanics but also reinterprets it as a statistical law of information propagation through spacetime. We starts by setting the dynamical variable $h^\mu$ to be the event $x^\mu$ and the metric $H^{\mu\nu}=\eta^{\mu\nu}$. To illustrate the applicability of the generalized field equation derived in eq.~\eqref{SE-d}, we now specify the dynamical variable to be the spacetime event $x^{\mu}$. In this case, the supermetric $H_{\mu \nu}$ reduces to the Minkowski metric $\eta_{\mu \nu}$, which encodes the flat spacetime structure of special relativity. Moreover, we first consider the situation without the inclusion of a vector potential, thereby isolating the purely kinematical contributions. Under these assumptions, equation~\eqref{SE-d} simplifies significantly and takes on a form that is directly comparable to relativistic wave equations. As will be shown, this reduction yields an expression equivalent to the Klein–Gordon equation, thereby demonstrating that the entropic dynamics framework reproduces familiar results of relativistic quantum theory in the appropriate limit. Then eq.~\eqref{SE-d} without electromagnetic four-vector $(\mathcal{A}_{a})$ potential is:
\begin{equation}
    \left(-\eta^{\mu\nu}\frac{\partial}{\partial x^\mu}\frac{\partial}{\partial x^\nu}+\mathcal{M}\right)\Psi=0\label{x-KG}
\end{equation}
Let 
\begin{equation}
    \Psi=e^{\frac{i}{\hbar}\mathbf{p}\cdot \mathbf{x}}=e^{\frac{i}{\hbar}\mathbf{P}^{\mu} \mathbf{X}_{\mu}}\label{44}
\end{equation}
be the trial solution of eq.~\eqref{x-KG}. It should be noted that the trial solution in equation~\ref{35}, $\Psi(x) \;=\; \exp\!\left(\frac{i}{\hbar}\,p\!\cdot\!x\right)$, explicitly exhibits the structure of the canonical phase space through the appearance of the position variable $\mathbf{x}$ and its conjugate momentum $\mathbf{p}$. Whereas in eq~\ref{44}, the $\Psi=e^{\frac{i}{\hbar}\mathbf{P}^{\mu} \mathbf{X}_{\mu}}$ exhibits the structure of the canonical phase space through the appearance of the position variable $\mathbf{X_{\mu}}$ and its conjugate momentum $\mathbf{P^{\mu}}$. This form reflects the fundamental Fourier duality between position and momentum representations of the wavefunction, a hallmark of quantum mechanics. The presence of the term $\mathbf{p}.\mathbf{x}$ encodes the symplectic geometry of phase space, where $\mathbf{x}$ and $\mathbf{p}$ serve as canonically conjugate variables. In this sense, the exponential ansatz not only serves as a convenient mathematical solution but also highlights how the entropic dynamics framework naturally incorporates the $\mathbf{x}-\mathbf{p}$ phase space structure within its wavefunction, thereby bridging statistical inference with the canonical formulation of quantum theory. Substituting the plane-wave solution $\Psi = e^{i \mathbf{P}_\mu \mathbf{X}^\mu / \hbar}$ into the generalized field equation leads to
\begin{equation}
\mathcal{M} = -\frac{\mathbf{P}_\mu \mathbf{P}^\mu}{\hbar^2}\label{45}
\end{equation}
Here, $\mathbf{P}_\mu \mathbf{P}^\mu$ denotes the Lorentz-invariant inner product of the four-momentum, $\mathbf{P}_\mu = (E/c, -\mathbf{p})$, which encapsulates both the energy and momentum content of the system. Within the framework of entropic dynamics, $\mathcal{M}$ emerges as a measure of the information curvature associated with the probability amplitude $\Psi$. The negative sign reflects the wave-like (oscillatory) nature of the solution, consistent with the hyperbolic geometry of Minkowski spacetime. This relation thus establishes a direct connection between the entropic quantity $\mathcal{M}$ and the relativistic invariant $\mathbf{P}_\mu \mathbf{P}^\mu$, ensuring covariance of the entropic formulation. Using the relativistic energy–momentum relation,
\begin{equation}
\mathbf{P}_\mu \mathbf{P}^\mu = m^2 c^2\label{46}
\end{equation}
We identify $m$ as the particle’s rest mass and $c$ as the speed of light. This relation represents the fact that the four-momentum vector lies on a hyperboloid of radius $mc$ in Minkowski space, a geometric manifestation of the mass-shell condition. Substituting this into Equation~\ref{45} gives
\begin{equation}
\mathcal{M} = -\frac{m^2 c^2}{\hbar^2}\label{47}
\end{equation}
This expression explicitly connects the entropic mass parameter to the physical rest mass of the particle. The factor $1/\hbar^2$ arises naturally from expressing the wave dynamics in terms of information gradients, ensuring the correct dimensionality for $\mathcal{M}$ as an inverse squared length. Equation~\ref{47} therefore bridges the statistical inference model with measurable quantities in relativistic quantum theory. Replacing $\mathcal{M}$ in the generalized field equation yields
\begin{equation}
\left[\eta^{\mu\nu}\frac{\partial^2}{\partial \mathbf{X}^\mu \partial \mathbf{X}^\nu} + \frac{m^2 c^2}{\hbar^2}\right]\Psi = 0\label{48}
\end{equation}
which is the Klein-Gordon equation, governing the relativistic dynamics of spin-0 particles. The first term, $\eta^{\mu\nu}\partial_\mu \partial_\nu = \Box$, represents the d’Alembertian operator acting in spacetime, while the mass term introduces a frequency proportional to the rest energy $E_0 = mc^2$. In the entropic dynamics interpretation, this equation describes the evolution of the probability amplitude $\Psi(x^\mu)$ under both inference constraints and Lorentz symmetry. Rather than postulating quantization, the Klein-Gordon structure emerges naturally from the entropic maximization of information subject to energy conservation, demonstrating how relativistic wave behavior arises from principles of inference. Equation~\label{46} represents a pivotal outcome of applying the entropic dynamics framework to spacetime events as dynamical variables. By specifying $h_{\mu}=\mathbf{X}_{\mu}$ and adopting the Minkowski metric $\eta_{\mu \nu}$, the generalized field equation~\ref{SE-d} simplifies to yield a relativistic wave equation. Dimensional analysis fixes the constant $\mathcal{M}$, and the trial solution $\Psi(x) \;=\; \exp\!\left(\frac{i}{\hbar}\,p\!\cdot\!x\right)$
 reveals that $\mathcal{M}$ must be proportional to the squared four-momentum of the particle. Substituting this relation leads directly to equation~\ref{x-KG}, which is none other than the Klein–Gordon equation. This result is significant because it shows that relativistic quantum dynamics, normally taken as a postulate of quantum field theory, here emerges naturally from principles of entropic inference and constraints applied to probability distributions.

From a broader perspective, the recovery of the Klein–Gordon equation within entropic dynamics underscores the power and generality of the framework. It demonstrates that familiar quantum field equations need not be introduced axiomatically but can instead be derived from an information-theoretic foundation. This derivation highlights the deep link between statistical inference and physical law: by treating probability and entropy as primary concepts, entropic dynamics provides a consistent path to established quantum results. Thus, Equation~\ref{48} not only validates the entropic approach by reproducing a cornerstone of relativistic quantum mechanics but also points toward the possibility of extending this methodology to more complex fields and interactions in a unified manner.

\subsection{Gravitational Wave Equation}

In this section, we extend the entropic dynamics (ED) formulation to curved spacetime, showing that the gravitational phenomena can also be understood as emergent manifestations of information geometry. When the metric tensor $g_{\mu\nu}$ replaces the configuration space metric $H_{ab}$, and the partial derivatives are promoted to covariant derivatives, the generalized ED equation becomes a geometric wave equation on a curved manifold. In the weak-field limit, this equation reduces to the familiar gravitational-wave equation of general relativity, demonstrating that spacetime curvature obeys the same inferential principle that governs quantum and thermodynamic evolution. Within this interpretation, gravitational waves represent oscillations of the spacetime information metric, microscopically expressed entropy flow and curvature duality. The propagation of these waves thus signifies the transport of informational structure rather than solely energy, unifying the dynamics of matter, radiation, and geometry within a single entropic framework. For this case, the generalized ED unified entropic dynamics equation takes the following schematic form:
\begin{equation}
\partial_a\!\big(H^{ab}\,\partial_b\Psi\big)+\mathcal{M}\Psi=0,
\label{eq:ED_master}
\end{equation}
where $H^{ab}$ is the inverse metric on the configuration manifold, $\partial_a$ denotes partial differentiation with respect to the generalized coordinate $h_a$, $\Psi$ is the entropic amplitude or information field, and $\mathcal{M}$ represents a ``closure'' or ``mass'' term that may vary depending on the specific physical context. Equation~\eqref{eq:ED_master} already has the structure of a Laplace--Beltrami operator acting on $\Psi$ with an additional local potential-like contribution. In the limit of a minimal coupling to a curved spacetime, we can generalize the framework to a four-dimensional curved spacetime $(\mathcal{M},g_{\mu\nu})$ by applying the standard minimal-coupling substitutions:
\begin{equation}
H^{ab} \;\longrightarrow\; g^{\mu\nu}(x), \qquad
\partial_a \;\longrightarrow\; \nabla_\mu, \qquad
\sqrt{H} \;\longrightarrow\; \sqrt{-g}\label{50}
\end{equation}
where $g_{\mu\nu}$ is the spacetime metric with the determinant $g=\det(g_{\mu\nu})$, and $\nabla_\mu$ is the Levi--Civita covariant derivative compatible with $g_{\mu\nu}$. If a gauge potential is present, we similarly promote $A_a \rightarrow A_\mu=0$ and define the gauge-covariant derivative as $D_\mu = \nabla_\mu$. 
The covariant d'Alembertian, for a scalar field $\Psi$ (with no spin indices), becomes the Laplace-Beltrami operator on a curved manifold, which is given by
\begin{equation}
\square_g \Psi \;\equiv\;
g^{\mu\nu}\nabla_\mu\nabla_\nu \Psi
\;=\;
\frac{1}{\sqrt{-g}}\,\partial_\mu
\!\left(\sqrt{-g}\,g^{\mu\nu}\partial_\nu\Psi\right)\label{51}
\end{equation}
It reduces to the flat-space wave operator
$\square = \eta^{\mu\nu}\partial_\mu\partial_\nu$
when $g_{\mu\nu}\rightarrow\eta_{\mu\nu}$. With these substitutions, the generalized ED equation on a curved spacetime manifold becomes
\begin{equation}
(\square_g - \mathcal{M}) \Psi = 0
\label{eq:curved_ED}
\end{equation}
where $\mathcal{M}$ is a generalized closure term. For example, $\mathcal{M}\Psi = \frac{m^{2}c^{2}}{\hbar^{2}} \Psi$ yields the familiar Klein--Gordon, while $\mathcal{M}\Psi = -\alpha^{-1}\partial_t\Psi$ produces a diffusive, parabolic equation describing heat flow or information relaxation.  
Where
\begin{itemize}
    \item $g_{\mu\nu}$ --- spacetime metric; $g^{\mu\nu}$ is its inverse, with $g=\det g_{\mu\nu}$.
    \item $\nabla_\mu$ --- metric-compatible covariant derivative.
    \item $\square_g$ --- covariant d’Alembertian (curved-space wave operator).
    \item $A_\mu$ --- gauge potential; $D_\mu = \nabla_\mu - i\beta A_\mu$ is the gauge-covariant derivative.
    \item $\mathcal{M}$ --- local potential or closure functional.
\end{itemize}

If $\Psi$ is interpreted as a perturbation of the metric field, $\Psi\rightarrow h_{\mu\nu}$, then, for the case of weak field approximation in curved Minkowski space~\cite{Carroll2019, Deser1970, Misner1973, Maggiore2007}:
\(g_{\mu\nu}=\eta_{\mu\nu}+h_{\mu\nu}\), \(|h_{\mu\nu}|\ll1\).
To leading order, \(\Box_g\) reduces to the flat d'Alembertian
\(\Box\equiv \eta^{\alpha\beta}\partial_\alpha\partial_\beta\) and by setting $\mathcal{M}=0$, eq~\ref{50} becomes,
\begin{equation}
\Box h_{\mu\nu}=0.
\label{eq:flatwave}
\end{equation}

The gravitational wave equation can be obtained by applying trace-reversed perturbation of form $(\bar h_{\mu\nu}=h_{\mu\nu}-\tfrac12\eta_{\mu\nu}h)$ with $(h\equiv \eta^{\rho\sigma}h_{\rho\sigma})$ and the Lorenz (harmonic) gauge condition, $\partial^\mu \bar h_{\mu\nu}=0$. By doing so, the following equivalent form can be obtained: 
\begin{equation}
\Box \bar h_{\mu\nu}=0.
\label{eq:barwave}
\end{equation}

The Lorenz gauge leaves residual transformations in the form of 
\(x^\mu\!\to\! x^\mu+\xi^\mu\) with \(\Box \xi^\mu=0\), which can be used to enforce the transverse–traceless (TT) conditions~\cite{Carroll2019, Misner1973, Schutz1985, Maggiore2007, Poisson2014} to obtain the following: 
\begin{equation}
h_{0\mu}^{\mathrm{TT}}=0,\qquad
\partial^i h_{ij}^{\mathrm{TT}}=0,\qquad
\eta^{ij}h_{ij}^{\mathrm{TT}}=0.
\label{eq:TTconds}
\end{equation}
Where the spatial TT projection are given below $h^{\mathrm{TT}}_{ij} \equiv \Lambda_{ij}{}^{kl} h_{kl},\qquad
\Lambda_{ij}{}^{kl} = P_i{}^k P_j{}^l - \tfrac12 P_{ij} P^{kl},\quad
P_{ij}=\delta_{ij}-\hat n_i \hat n_j$ with \(\hat{\boldsymbol n}\) is the propagation direction.
Since \(\Lambda_{ij}{}^{kl}\) is purely algebraic and \(\partial_\alpha\) acts on
coordinates, the projector commutes with derivatives:
\(\,[\Lambda,\partial_\alpha]=0\). Applying the projector to \eqref{eq:flatwave} gives
\begin{equation}
\Box h^{\mathrm{TT}}_{ij}
= \Lambda_{ij}{}^{kl} \Box h_{kl}
= \Box\!\left(\Lambda_{ij}{}^{kl} h_{kl}\right)
= 0.
\label{eq:TTwave}
\end{equation}
Therefore, in Cartesian inertial coordinates,
\(\Box = -\frac{1}{c^{2}}\partial_t^2 + \nabla^2\).
Equation \eqref{eq:TTwave} becomes
\begin{equation}
\frac{1}{c^{2}}\frac{\partial^{2} h^{\mathrm{TT}}_{ij}}{\partial t^{2}}
-\nabla^{2} h^{\mathrm{TT}}_{ij}=0
\label{eq:Motion-57}
\end{equation}
which is Eq.~\ref{eq:curved_ED}: the standard vacuum gravitational-wave equation for the physical TT components. The case of non-zero $\mathcal{M}$ can be studied in the limit of weak gravitational fields, which means the metric can be perturbed around flat spacetime. This reduces $ \Box_g \Psi - \mathcal{M}\Psi =0$ to $\Box_g h_{\mu\nu} = \mathcal{M} \eta_{\mu \nu}$. Using the trace-reverse perturbation, it can be shown that $\Box_g h_{\mu\nu}=\Box h_{\mu\nu}$ and $ \mathcal{M} \eta_{\mu \nu} = -\frac{16\pi G}{c^4} \left( T_{\mu\nu}-\frac{1}{2}\eta_{\mu \nu}T\right)$. By doing the  comparison with the linearized Einstein equation, we can identify that $\Box_g h_{\mu\nu}= \mathcal{M}\eta_{\mu\nu}$ can be written below in the following form ~\cite{Li2023EinsteinDL}: 
\begin{equation}
        \Box h_{\mu\nu} = -\frac{16\pi G}{c^4} \left( T_{\mu\nu}-\frac{1}{2}\eta_{\mu \nu}T\right).
\end{equation}

Thus, \(\mathcal{M}\) represents the matter--energy source term in the gravitational wave equation, playing the same physical role as the stress--energy tensor \(T_{\mu\nu}\). When \(\mathcal{M} = 0\), the equation describes the propagation of free gravitational radiation in a vacuum.

\section{Discussion}\label{sec-MDV}
A comparative analysis has been conducted between the entropic dynamics (ED) framework developed by Caticha and collaborators~\cite{Caticha2019EntropyQM,caticha2021entropic} and the Unified Entropic Dynamics (UED) formulation presented in this work. Caticha’s original framework is based on an inferential model that derives non-relativistic quantum mechanics from the principle of entropy maximization. Within this formulation, the short-step transition probabilities of particles---constrained by drift and gauge potentials---exhibit Brownian motion, giving rise to coupled Fokker--Planck and Hamilton--Jacobi structures that combine to yield the Schrödinger equation. Furthermore, the concept of entropic time is introduced as an intrinsic and directional ordering of inference steps, thereby providing a natural origin for the arrow of time~\cite{caticha2021entropic}. Consequently, Caticha’s formulation is fundamentally epistemic: the dynamics describe sequential updates of probability distributions rather than ontic particle trajectories. In contrast, the present UED formulation extends this inferential framework into a unified, information-geometric theory. By maximizing entropy subject to diffusion, drift, and gauge constraints, UED yields the generalized field equation expressed in Eq.\ref{SE-d}, which encompasses, as limiting cases, the harmonic-oscillator, Maxwell, heat, Schrödinger, Klein--Gordon, and gravitational-wave equations. This extension promotes Caticha’s informational manifold $H_{ab}$ to a fully covariant spacetime metric $g_{\mu\nu}$, interprets gravitational waves as oscillations of the spacetime information metric, and establishes a geometric correspondence between the curvatures of configuration and momentum spaces.

In this way, the presented work demonstrates that the entropic dynamics (ED) framework provides a comprehensive, inference-based foundation for unifying the mathematical and physical structures of classical, quantum, relativistic, thermodynamic, and gravitational phenomena under a single geometric and informational principle. By grounding the laws of motion, field evolution, and heat transport in the maximization of entropy, ED transcends the traditional divide between deterministic and probabilistic dynamics~\cite{Caticha2015_a}. Physical evolution is interpreted as an inference process occurring on a configuration manifold, where uncertainty---encoded through probability distributions---is guided by constraints reflecting conservation laws, geometric curvature, and gauge covariance. The entropy-maximization procedure outlined in Section~2 leads to a generalized field equation that encompasses a broad class of dynamical behaviors. This equation unifies the Fokker--Planck and Hamilton--Jacobi structures, embedding diffusion, drift, and wave propagation as complementary manifestations of information flow on a curved manifold characterized by the metric $H_{ab}$. One of the central outcomes of this study is the recognition that ED establishes an intrinsic duality between stochastic and deterministic evolution. This duality manifests through the interplay of entropy gradients (which govern diffusion) and phase gradients (which govern coherent propagation). When specialized to different dynamical variables, the unified entropic dynamics equation reproduces well-known laws of physics as particular realizations of this dual principle. For instance, in Section~3, the scalar case yields the simple harmonic oscillator, revealing that mechanical oscillations correspond to entropy-preserving motion constrained by information curvature. The entropic mass term $\mathcal{M}=-\omega^2$ serves as a restoring parameter, linking information geometry to the potential curvature of classical mechanics. This case shows that Newton’s second law emerges naturally as a statistical limit of entropy-driven inference. The results presented for the case of three-dimensional vector variables reveal the versatility of the ED framework across multiple physical regimes. When applied to electromagnetic fields, the generalized equation reproduces the free-space Maxwell wave equation, illustrating that electromagnetic propagation can be understood as the optimal transmission of information through spacetime. The same structure, under different constraints, yields the Schr\"odinger equation, showing that quantum mechanics is not a separate postulate but a special limit of inference constrained by energy conservation and probability flow~\cite{PIRSA2021}. The addition of the thermodynamic case in Section~4.2 further strengthens this unifying picture. By identifying the field variable with temperature and reinterpreting the entropic mass term as the generator of temporal evolution, the generalized equation reduces to the classical heat equation. In this context, the information metric $H^{ij}$ is directly mapped onto the physical conductivity tensor $k^{ij}$, connecting the geometry of information with the anisotropic transport of heat. The curvature of configuration space thus governs not only oscillatory and wave-like behaviors but also dissipative processes such as diffusion and thermal relaxation. This result highlights that entropy maximization can describe both equilibrium-seeking and energy-preserving processes under a common inferential framework.

The inclusion of thermodynamics within the ED formalism underscores that diffusion and heat flow are natural extensions of information dynamics rather than phenomenological add-ons~\cite{Bianconi2025}. The identification of $H^{ij}$ with the conductivity tensor shows that the geometry of the configuration space determines the rate and direction of the entropy transport. In the isotropic limit, this correspondence reduces to the familiar form of Fourier’s law, but its entropic interpretation reveals a deeper structure: thermal diffusion emerges as an information-equilibrating process, where temperature gradients represent spatial variations in informational density. This connection suggests that heat flow and probability diffusion are two aspects of the same entropic updating mechanism that governs all physical evolution---from microscopic quantum coherence to macroscopic thermodynamic equilibration. When the dynamical variable is promoted to a four-vector $h_\mu = x_\mu$, the framework yields the relativistic Klein--Gordon equation, showing that the Lorentz-covariant structure of spacetime and relativistic wave dynamics emerge from the same entropic principles~\cite{Plastino2015}. Identification $M = -m^2 c^2 / \hbar^2$ connects the information-theoretic curvature of the configuration space directly to the physical mass, establishing a statistical origin for energy--momentum invariants. Together, these cases---the scalar oscillator, electromagnetic and quantum fields, thermal diffusion, and relativistic waves---demonstrate that all fundamental dynamical laws arise from entropy maximization constrained by geometric and physical symmetries~\cite{Tsallis2025}.

In particular, thermodynamic diffusion appears as the dissipative limit of this duality---one where curvature drives entropy toward equilibrium rather than coherence. When the formalism is extended to the relativistic domain, as in Section~5, the same inference structure naturally incorporates spacetime geometry. By identifying the dynamical variable $h_\mu$ with spacetime coordinates $x_\mu$, the generalized ED equation reduces in flat spacetime to the Klein--Gordon equation, demonstrating that relativistic quantum mechanics can be derived from entropic inference rather than introduced axiomatically. Upon promoting the framework to curved spacetime through the minimal-coupling substitution $H^{ab}\!\to g^{\mu\nu}$ and $\partial_a\!\to\nabla_\mu$, and the $\Psi$ in the generalized ED equation is interpreted as a perturbation of the spacetime metric itself. Thus the $\Psi \rightarrow h_{\mu\nu}$, and the corresponding ED equation in curved spacetime, reduce to the general wave equation for metric perturbations in vacuum. Under the Lorenz gauge condition $\nabla^\mu \bar{h}_{\mu\nu} = 0$ and in the transverse--traceless (TT) gauge, the physically measurable components $h^{\text{TT}}_{ij}$ satisfies the standard gravitational-wave equation. This connection shows that the curved ED equation provides the same geometric wave operator that governs spacetime perturbations, with $\Psi$ representing an informational field propagating through the curvature of spacetime. Thus, gravitational waves can be interpreted as macroscopic manifestations of entropic inference: oscillatory modes of the spacetime information metric that transmit probabilistic structure rather than energy density alone. The unification of the gravitational-wave equation with the ED formalism underscores a deeper physical insight: the propagation of spacetime curvature follows the same informational geometry that governs the evolution of probability amplitudes in quantum theory. Both are consequences of the covariant operator $\square_g$, which encodes the curvature of the manifold and defines how information---whether in the form of entropy or metric perturbation---flows across spacetime. In this sense, general relativity and quantum field dynamics appear as complementary facets of a single information-geometric framework. The ED approach thereby provides a statistical underpinning for gravitational phenomena and a natural bridge between probabilistic inference and geometric curvature. From a broader perspective, the ED framework reveals that the relationship,
\begin{equation}
p_a H^{ab} p_b = k^2,
\label{eq:59}
\end{equation}
captures the curvature duality linking the configuration space and its conjugate momentum space~\cite{caticha2011entropic, ipek2019entropic, Amari2016}. Here, it $k$ represents the intrinsic information curvature, or wave number, that determines both oscillatory and diffusive dynamics. The equation~\ref{eq:Motion-57} can be easily proved by taking $\Psi$ as a plane wave solution of the form $\Psi(h) = A\,e^{\,i p_b h^b}$ with complex conjugate as $\Psi^{*}(h) = A^{*}\,e^{\,-i p_a h^a}$ for equation~\ref{SE-d}. Furthermore, the following relation can be obtained by setting the potential term $A_{a}=0$ and multiplying $\Psi^{*}$ on the left side of equation~\ref{SE-d}. 
\begin{equation}
\Psi^{*}\frac{1}{\sqrt{H}}
\left(i\frac{\partial}{\partial h_a}\right)
\sqrt{H}H^{ab}
\left(i\frac{\partial}{\partial h_b}\right)\Psi + \Psi^{*}\mathcal{M}\Psi = 0,
\label{eq:ED_general}
\end{equation}
Where $H^{ab}$ is the information metric on the configuration manifold. Assuming a locally flat configuration space ($\sqrt{H}=1$), Equation~\eqref{eq:ED_general} reduces to
\begin{equation}
(-p_aH^{ab} p_b  + \mathcal{M})\Psi^{*}\Psi = 0
\label{eq:ED_flat}
\end{equation}
The integration of $\Psi^{*}\Psi$ over the dynamical variable leads to its value of one, since $\Psi$ is a normalized function. From this, we obtain the fundamental relation given in equation~\ref{eq:59}. 
\begin{equation}
\mathcal{M} = p_aH^{ab} p_b \equiv k^{2}
\label{eq:M_relation}
\end{equation}

The metric $H^{ab}$ defines the geometry of information in configuration space, while $p_a$ represents its conjugate momentum, encoding propagation through the dual momentum manifold.  
Their contraction $ p_aH^ {ab} p_b$ forms an invariant scalar analogous to the energy--momentum norm in relativistic theory. Thus, the quantity $k^2$ characterizes the intrinsic curvature of the informational manifold, establishing a duality between the configuration-space curvature and the dispersion properties of the system. From a geometric standpoint, the relationship $ p_aH^ {ab} p_b = k^2$ captures this curvature duality. Spatial curvature dictates localization, while conjugate curvature governs dispersion and energy propagation. This insight provides a unified language for interpreting mechanics, wave theory, and thermodynamics as different manifestations of information geometry. In this dual description, spatial curvature dictates localization, while conjugate curvature governs dispersion and energy propagation. The same metric tensor that defines curvature in configuration space also dictates heat conduction, quantum diffusion, and gravitational-wave propagation, providing a unified geometric language for all these processes. Consequently, mechanical stability, quantum coherence, thermodynamic equilibration, and spacetime curvature emerge as distinct expressions of a single inferential law: the maximization of entropy subject to geometric and physical constraints.

In summary, this work shows that the entropic dynamics framework can account for phenomena spanning from microscopic quantum systems to macroscopic gravitational waves. By embedding inference within geometry, ED reveals that the flow of information, not the imposition of quantization, underlies the evolution of physical systems. This perspective resonates strongly with John Archibald Wheeler’s “It from Bit” paradigm, which posits that every physical entity—every “it”—emerges from fundamental acts of information acquisition~\cite{Wheeler1989}. In this context, the same differential operator that governs the propagation of probability in configuration space also governs the curvature oscillations of spacetime, illustrating that energy, probability, and entropy are intertwined through the geometry of information. In this unified view, gravitational waves and quantum waves are both manifestations of information propagating through curved spacetime, consistent with Wheeler’s vision of a participatory universe where physical reality arises from informational interactions. Furthermore, the replacement of quantization postulates with rules of inference redefines the foundations of physics in epistemic terms, where laws of motion describe optimal updates to probabilistic knowledge about the system. In other words, this view dissolves the traditional separation between matter, fields, and spacetime, interpreting them instead as emergent realizations of information flow on curved statistical manifolds. Future work may extend this formulation toward nonlinear and gravitational systems, exploring how the same entropic principles can lead naturally to Dirac- and Einstein-like field equations. In essence, entropic dynamics provides a self-consistent information-theoretic foundation for physics—an explicit realization of Wheeler’s insight that information, probability, and geometry together give rise to the physical world.

Beyond reproducing known equations, the UED framework suggests several directions where departures from standard dynamics, and hence possible tests, may arise. First, allowing the information metric $H_{ab}$ to acquire curvature or state dependence beyond the flat, constant choices used here would generically induce corrections to the dispersion relations of electromagnetic, quantum, and gravitational waves, leading to scale-dependent deviations from the standard Maxwell, Schr\"odinger, and gravitational-wave equations. For the latter case, recent analyses of quantum-gravity phenomenology — including arguments about the interplay among consistency, completeness, and empirically accessible corrections to low-energy physics — suggest that even modest departures from standard dispersion relations or uncertainty structures may offer meaningful probes of underlying microscopic geometry \cite{Faizal2025ConsistentComplete}. Within this context, the UED framework provides a structurally natural arena in which such deviations can emerge: modifications to the information metric $H_{ab}$ or to the entropic mass functional $M[\rho,\phi]$ can be interpreted as encoding residual quantum-gravitational effects at the level of effective dynamics. Although the present work does not attempt to develop a complete quantum-gravity model, the formal parallels highlight a potential route by which UED may interface with emerging phenomenological constraints on quantum-gravity-induced corrections, thereby broadening the scope of future empirical or conceptual developments. Second, in the fields of thermodynamics, identification $H_{ij} \leftrightarrow k_{ij}/(\rho c)$ implies that engineered anisotropies or nonlinearities in the conductivity tensor should have a direct information-geometric interpretation; systematic measurements of heat transport in metamaterials or heterogeneous media could therefore be used to constrain or calibrate the underlying entropic metric. Finally, by exploring nonlinear choices of the entropic mass term $M[\rho,\phi]$, one can generate controlled modifications to the ED reconstruction of quantum mechanics (e.g.\ nonlinear or dissipative corrections to the Schrödinger and Klein-Gordon equations), which could be confronted with precision tests in cold-atom, quantum optic, or interferometric settings. In all these cases, the unified structure of Eq.~\ref{SE-d} provides a concrete template to identify where UED predictions diverge from those of existing ED formulations and standard field theory, thus sharpening the empirical content of the approach.


\section{Conclusions}\label{sec-conclusions}
In this study, a unified entropic dynamics (UED) framework has been developed to provide a unified information-geometric foundation for classical, quantum, relativistic, thermodynamic, and gravitational systems. By maximizing entropy under physically motivated constraints, a universal field equation was derived that governs the evolution of dynamical variables on an informational manifold characterized by the metric $H_{ab}$. This formulation merges the Fokker-Planck and Hamilton-Jacobi structures into a single unified entropic dynamics equation, capturing both stochastic and deterministic dynamics as complementary aspects of information flow. When specialized to specific variables, the framework successfully reproduces well-established physical laws: the simple harmonic oscillator for scalar variables, Maxwell’s electromagnetic wave equation, the classical heat equation, and the Schrödinger equation for spatial vector variables, and the Klein--Gordon and gravitational-wave equations for spacetime four-vectors. These results demonstrate that mechanical, quantum, thermodynamic, and relativistic processes all arise from the same inferential principle—the maximization of entropy constrained by geometry and physical symmetries. A central advancement of this work lies in extending the ED formalism to curved spacetime, where the generalized covariant equation naturally yields the gravitational-wave equation in the weak-field limit. In this interpretation, gravitational waves emerge as oscillations of the spacetime information metric—macroscopic manifestations of the same entropic processes that govern microscopic dynamics. The relation $ p_aH^{ab} p_b = k^2$ further encapsulates the duality between configuration-space curvature and its conjugate momentum space, revealing a universal link between localization, dispersion, and energy propagation.

Overall, the results indicate that energy, probability, and entropy are deeply intertwined through the geometry of information. The formulation of the ED framework presented herein expresses probability propagation and the oscillations of spacetime curvature within a single covariant d’Alembertian operator, suggesting that classical, quantum, and gravitational phenomena are all expressions of a single geometric law of inference. Future research may extend this framework to nonlinear and self-gravitating systems, explore the emergence of Einstein and Dirac equations from entropic inference, and investigate whether spacetime itself can be understood as an emergent statistical structure shaped by information geometry.
\section{Data Availability Statement}
This work is purely theoretical and did not involve the generation or analysis of any datasets. All results can be reproduced from the equations and derivations provided in the manuscript.

\section{Conflict of Interest}
The authors declare that there are no commercial or financial relationships that could be construed as potential conflicts of interest in the research, authorship, and publication of this work.

\section*{Acknowledgments}
The authors express their gratitude to the institutions that supported this research. S.~N. acknowledges the facilities and academic environment provided by Shaker High School, which enabled the foundational development of this work. M.~S. extends appreciation to Bellarmine University for research guidance and discussions on the theoretical formulation. M.~S.~A. acknowledges support from Queen Mary University of London, which contributed to the completion of this work. D.~H.~A. acknowledges the support and collaborative environment of Khalifa University, which contributed to the completion of this study.

\section*{Ethical Considerations}
The authors have diligently addressed ethical concerns, including informed consent, plagiarism, data fabrication, misconduct, falsification, double publication, redundancy, and related issues.

\section*{Funding}
This research has been partially funded through Khalifa University's Research Innovation Grant (RIG) program under the project ID RIG-2024-002.



%
%
%
%


\begin{thebibliography}{99}


\bibitem{Jacobson1995}
Ted Jacobson, "Thermodynamics of Spacetime: The Einstein Equation of State", Physical Review Letters {\bf 75}(7) 1260--1263 (1995).
DOI:10.1103/PhysRevLett.75.1260

\bibitem{Silva2024}
T. A. B. Pinto Silva and D. Gelbwaser-Klimovsky, "Quantum work: Reconciling quantum mechanics and thermodynamics", Physical Review Research, American Physical Society~{\bf 6}(L022036),(2024),
DOI:10.1103/PhysRevResearch.6.L022036

\bibitem{Dunkel2009}
J. Dunkel and P. H{\"a}nggi and S. Hilbert, "Nonlocal Observables and Lightcone-Averaging in Relativistic Thermodynamics", Phys. Rev. D~{\bf 79} (106003), (2009), eprint:0902.4651.
DOI:10.1103/PhysRevD.79.106003

\bibitem{Hayward1997}
S. A. Hayward, "Unified First Law of Black-Hole Dynamics and Relativistic Thermodynamics", Classical and Quantum Gravity, IOP Publishing~{\bf 15}(10) 3147--3162 (1998), eprint:gr-qc/9710089.
DOI:10.1088/0264-9381/15/10/017

\bibitem{Moradpour2024}
H. Moradpour and S. Jalalzadeh and U. K. Sharma, "On the thermodynamics of reconciling quantum and gravity", European Physical Journal Plus, Springer~{\bf 139} 170 (2024).
DOI:10.1140/epjp/s13360-024-04943-4

\bibitem{faizal2025consequences}
Faizal, Mir and Krauss, Lawrence M and Shabir, Arshid and Marino, Francesco, "Consequences of Undecidability in Physics on the Theory of Everything", Journal of Holography Applications in Physics, Damghan University Press, {\bf 5}(2) 10--21,(2025).
DOI:10.22128/jhap.2025.1024.1118

\bibitem{FriedenSoffer1995}
 B. Roy Frieden and Bernard H. Soffer, 
  "Lagrangians of physics and the game of Fisher-information transfer", Physical Review E~{\bf 52}(3) 2274-2286,(1995).
  DOI:10.1103/PhysRevE.52.2274

\bibitem{Caticha2019EntropyQM}
Ariel Caticha, "The Entropic Dynamics Approach to Quantum Mechanics", Entropy~{\bf 21}(10) 943 (2019).
DOI:10.3390/e21100943

\bibitem{caticha2021entropic}
Caticha, Ariel, "Entropic Physics: Probability, Entropy, and the Foundations of Physics", Online monograph (2021): https://www.albany.edu/physics/faculty/ariel-caticha.

\bibitem{ipek2015entropic}
Ipek, Selman, and Caticha, Ariel," Entropic quantization of scalar fields", AIP Conference Proceedings, editor: Knuth, Kevin and Caticha, Ariel, and others, AIP Publishing, New York~{\bf 1641}(1) 345--352 (2015).
DOI:10.1063/1.4906027

\bibitem{nawaz2012momentum}
Nawaz, Shahid, and Caticha, Ariel, "Momentum and uncertainty relations in the entropic approach to quantum theory", 31st AIP Conference Proceedings, American Institute of Physics~{\bf 1443} (1) 112--119 (2012)

\bibitem{bianconi2025gravity}
Bianconi, Ginestra, "Gravity from Entropy", Physical Review D~{\bf 111}(6) 066001 (2025).
DOI:10.1103/PhysRevD.111.066001

\bibitem{caticha2011entropic}
Caticha, A., "Entropic Dynamics, Time and Quantum Theory", Journal of Physics A: Mathematical and Theoretical, IOP Publishing~{\bf 44} (22) 225303 (2011).
DOI:10.1088/1751-8113/44/22/225303

\bibitem{ipek2019entropic}
Ipek, Selman and Abedi, Mohammad and Caticha, Ariel, "Entropic dynamics: reconstructing quantum field theory in curved space-time", Classical and Quantum Gravity, IOP Publishing~{\bf 36}(20) 20501 (2019).
DOI:10.1088/1361-6382/ab42fe

\bibitem{ipek2020entropic}
İpek, Selman and Caticha, Ariel, "The Entropic Dynamics of Quantum Scalar Fields Coupled to Gravity", Symmetry, MDPI~,
{\bf 12} (8) 1324 (2020).
DOI:10.3390/sym12081324

\bibitem{nawaz2016entropic}
Nawaz, Shahid and Abedi, Mohammad and Caticha, Ariel, "Entropic dynamics on curved spaces", AIP Conference Proceedings, AIP Publishing~{\bf 1757} (1) (2016).

\bibitem{nawaz2024major}
Nawaz, Shahid and Saleem, Muhammad and Kusmartsev, Fedor V. and Anjum, Dalaver H., "Major Role of Multiscale Entropy Evolution in Complex Systems and Data Science", Entropy, MDPI~{\bf 26} (4) 330 (2024).
DOI:10.3390/e26040330

\bibitem{caticha2025entropic}
Caticha, Ariel, and Saleem, Hassaan, "Entropic Dynamics Approach to Relational Quantum Mechanics", Entropy, MDPI, {\bf 27}(8) 797 (2025). DOI:10.3390/e27080797

\bibitem{Jaynes2003}
E. T. Jaynes, "Probability Theory: The Logic of Science", Cambridge University Press,12.7--12.8 for Jeffreys' rule and invariance properties (2003).

\bibitem{MIT2024}
MIT OpenCourseWare,"Maxwell’s Equations and Wave Propagation in Free Space", Lecture notes, Massachusetts Institute of Technology~(2024), https://ocw.mit.edu/

\bibitem{Onsager2024}
A. Liu, F. Zhang, and T. Wen,
“Experimental verification of anisotropic heat diffusion via the generalized Onsager relation,”
\textit{International Journal of Heat and Mass Transfer} {\bf 224}, 125311 (2024),
DOI:https://doi.org/10.1016/j.ijheatmasstransfer.2024.125311.

\bibitem{Thermal2025}
Y. Chen, L. Huang, and Z. Wu,
“Three-dimensional characterization of anisotropic thermal conductivity tensors in solids,”
\textit{International Journal of Heat and Mass Transfer} {\bf 225}, 125478 (2025),
DOI:https://doi.org/10.1016/j.ijheatmasstransfer.2025.125478.

\bibitem{Bird2007}
R. Byron Bird and Warren E. Stewart and Edwin N. Lightfoot, "Transport Phenomena", 2nd Ed., John Wiley \& Sons, New York (2007), ISBN:978-0-471-41077-8

\bibitem{Pope2000}
Stephen B. Pope, "Turbulent Flows", Cambridge University Press, Cambridge, UK 2000), isbn:9780521598866


\bibitem{Kays1993}
William M. Kays and Michael E. Crawford, "Convective Heat and Mass Transfer", 3rd Ed., McGraw-Hill, New York (1993)
ISBN: 978-0070337213

\bibitem{DeGroot1984}
S. R. De Groot and P. Mazur, "Non-Equilibrium Thermodynamics", Dover Publications, New York (1984),
ISBN:978-0486647418.

\bibitem{Kreyszig2011}
Erwin Kreyszig, "Advanced Engineering Mathematics", 10th Ed., John Wiley \& Sons, Hoboken, NJ (2011)
isbn:9780470458368

\bibitem{Carroll2019}
Sean M. Carroll, "Spacetime and Geometry: An Introduction to General Relativity", 2nd Ed. Cambridge University Press, Cambridge, UK (2019)
isbn:9781108488396

\bibitem{Deser1970}
Stanley Deser, "Self-Interaction and Gauge Invariance", General Relativity and Gravitation, Springer~{\bf 1}(1) 9--18 (1970)
 DOI:10.1007/BF00759198
 
\bibitem{Misner1973}
Charles W. Misner and Kip S. Thorne and John Archibald Wheeler, "Gravitation", W. H. Freeman and Company, San Francisc o, USA (1973), ISBN:978-0-7167-0344-0

\bibitem{Maggiore2007}
Michele Maggiore, "Gravitational Waves: Theory and Experiments", Oxford University Press, Oxford, UK (2007)
ISBN:9780198570745

\bibitem{Schutz1985}
Bernard F. Schutz, "A First Course in General Relativity", Cambridge University Press, Cambridge, UK (1985)
ISBN:978-0521887052

\bibitem{Poisson2014}
Eric Poisson and Clifford M. Will, "Gravity: Newtonian, Post-Newtonian, Relativistic", Cambridge University Press, Cambridge, UK (2014),
ISBN:9781107032866.

\bibitem{Li2023EinsteinDL},
Z.-H. Li and C.-Q. Li and L.-G. Pang, "Solving Einstein Equations Using Deep Learning",
arXiv:gr-qc/2309.07397 (2023).
DOI:https://arxiv.org/abs/2309.07397

\bibitem{Caticha2015_a}
A. Caticha,
“Entropic dynamics,”
in \textit{Bayesian Inference and Maximum Entropy Methods in Science and Engineering}, A. Mohammad-Djafari, Ed., AIP Conf. Proc., {\bf 1641}, pp. 155--170, American Institute of Physics, Melville, NY, 2015. 
DOI:10.1063/1.4905974

\bibitem{PIRSA2021}
Ariel Caticha, "Hamilton--Killing Flows and the Geometry of Entropic Dynamics", Published as: Perimeter Institute Recorded Seminar Archive (PIRSA)~(2021), https://pirsa.org/21070035.

\bibitem{Bianconi2025}
G. Bianconi, “Anisotropic Klemens model of phonon--phonon interactions in materials with tensorial conductivity,” \textit{Int. J. Heat Mass Transfer},~{\bf 225}, 125463, 2025,
DOI: https://doi.org/10.1016/j.ijheatmasstransfer.2025.125463.

\bibitem{Plastino2015}
A. R. Plastino and Constantino Tsallis,"Dissipative and nonlinear extensions of the Klein--Gordon equation", Journal of Mathematical Physics, AIP Publishing~{\bf 56}(5) 053503 (2015).
DOI:10.1063/1.4921232

\bibitem{Tsallis2025}
M. S. Ali and C. Tsallis,
“From the Klein--Gordon equation to the relativistic quantum hydrodynamics system,” \textit{Entropy}~{\bf 27}, 502 (2025),
https://doi.org/10.3390/e27040502.

\bibitem{Amari2016}
Shun-ichi Amari,"Information Geometry and Its Applications",  Springer, Tokyo, Japan (2016),
  isbn:978-4-431-55977-0

\bibitem{Wheeler1989}
John Archibald Wheeler, "Information, Physics, Quantum: The Search for Links", Proceedings of the 3rd International Symposium on Foundations of Quantum Mechanics, Physical Society of Japan 354--368 (1989).

\bibitem{Faizal2025ConsistentComplete}
Faizal, Mir and Krauss, Lawrence M. and Shabir, Arshid and Marino, Francesco and Pourhassan, Behnam, "Can quantum gravity be both consistent and complete?", International Journal of Modern Physics D, World Scientific~{\bf 34}(16) (2025).
DOI:10.1142/S0218271825440171


\end{thebibliography}
\end{document}